\documentclass[a4paper,fleqn,usenatbib]{mnras}

\usepackage[T1]{fontenc}
\usepackage{ae,aecompl}
\usepackage{graphicx}
\usepackage{amsmath}
\usepackage{amssymb}

\newcommand{\Msun}{\,\mathrm{M}_{\odot}}

\newcommand{\pc}{\,\mathrm{pc}}
\newcommand{\kpc}{\,\mathrm{kpc}}
\newcommand{\yr}{\,\mathrm{yr}}
\newcommand{\Myr}{\,\mathrm{Myr}}
\newcommand{\Gyr}{\,\mathrm{Gyr}}
\newcommand{\kms}{\,\mathrm{km\,s}^{-1}}

\newcommand{\Nbody}{$N$-body }
\newcommand{\Texp}{T_\mathrm{exp}}
\newcommand{\Tdyn}{T_\mathrm{dyn}}

\newcommand{\percent}{\,per\,cent}

\newcommand{\Rt}{R_\mathrm{t}}
\newcommand{\Ra}{R_\mathrm{a}}
\newcommand{\Rp}{R_\mathrm{p}}
\newcommand{\rs}{r_\mathrm{s}}
\newcommand{\Rh}{R_\mathrm{h}}
\newcommand{\RG}{R_\mathrm{G}}
\newcommand{\RGC}{R_\mathrm{GC}}
\newcommand{\Rc}{R_\mathrm{c}}

\newcommand{\Mc}{M_\mathrm{c}}

\newcommand{\phiG}{\phi_\mathrm{G}}

\newcommand{\Vc}{V_\mathrm{c}}
\newcommand{\Vinfty}{V_\infty}

\newcommand{\secref}[1]{Section~\ref{#1}}
\newcommand{\figref}[1]{Fig.~\ref{#1}}
\newcommand{\tabref}[1]{Table~\ref{#1}}
\newcommand{\equref}[1]{equation~\eqref{#1}}

\defcitealias{Khalaj2015}{KB15}
\defcitealias{D'Ercole2008}{DE08}

\title[significant mass-loss in the GCs of the Fornax dSph]{Limits on the significant mass-loss scenario based on the globular clusters of the Fornax dwarf spheroidal galaxy}

\author[P. Khalaj and H. Baumgardt]{P. Khalaj\thanks{E-mail: pouria.khalaj@uqconnect.edu.au} and H. Baumgardt\thanks{E-mail: h.baumgardt@uq.edu}\\
School of Mathematics and Physics, University of Queensland, St. Lucia, QLD 4072, Australia}

\date{Accepted 2015 December 14. Received YYY; in original form ZZZ}
\pubyear{2016}

\begin{document}
\label{firstpage}
\pagerange{\pageref{firstpage}--\pageref{lastpage}}
\maketitle

\begin{abstract}
Many of the scenarios proposed to explain the origin of chemically peculiar stars in globular clusters (GCs) require significant mass-loss ($\ge95\%$) to explain the observed fraction of such stars. In the GCs of the Fornax dwarf galaxy significant mass-loss could be a problem. \citet{Larsen2012} showed that there is a large ratio of GCs to metal-poor field stars in Fornax and about $20-25\%$ of all the stars with ${\rm [Fe/H]}<-2$ belong to the four metal-poor GCs. This imposes an upper limit of $\sim80\%$ mass-loss that could have happened in Fornax GCs. In this paper, we propose a solution to this problem by suggesting that stars can leave the Fornax galaxy. We use a series of \Nbody simulations, to determine the limit of mass-loss from Fornax as a function of the initial orbital radii of GCs and the speed with which stars leave Fornax GCs. We consider a set of cored and cuspy density profiles for Fornax. Our results show that with a cuspy model for Fornax, the fraction of stars which leave the galaxy, can be as high as $\sim 90\%$, when the initial orbital radii of GCs are $R=2-3\kpc$ and the initial speed of stars is $v>20\kms$. We show that such large velocities can be achieved by gas expulsion induced mass-loss but not stellar evolution induced mass-loss. Our results imply that one cannot interpret the metallicity distribution of Fornax field stars as evidence against significant mass-loss in Fornax GCs, if mass-loss is due to gas expulsion.
\end{abstract}



\section{Introduction}
Almost all GCs in the Milky Way and other galaxies such as the Fornax dwarf spheroidal galaxy (dSph) show light element abundance anomalies and splits in their colour-magnitude diagram which can be attributed to the existence of more than one stellar population in these GCs. Examples of such GCs include $\omega$\,Cen \citep{Rey2004}, NGC2808 \citep{Piotto2007, Milone2015}, GCs in Fornax \citep{D'Antona2013} and M80 \citep{Carretta2015}.

\par There are several scenarios which have been proposed to explain the origin of chemically peculiar stars in GCs, e.g. fast-rotating massive stars (FRMS; $20-120\Msun$; \citealt{Prantzos2006, Decressin2007a, Decressin2007b, Krause2013}, asymptotic giant branch stars (AGB; $4-9\Msun$; \citealt{Ventura2001, D'Ercole2008}), intermediate-mass and massive binaries ($4-100\Msun$; \citealt{deMink2009}) and early disc accretion in low-mass pre-main-sequence stars (polluter mass $M>10\Msun$; \citealt{Bastian2013}). Among these scenarios, FRMS and the AGB scenario are challenged by the so-called mass-budget problem. According to these scenarios, the ejecta of primordial stars which later form chemically peculiar stars make up between $4$ and $9${\percent} of the initial stellar mass \citep{deMink2009}, assuming a standard \cite{Kroupa2001} mass function for primordial stars. Unless the initial mass-function of primordial stars was top-heavy, this is not sufficient for the formation of a large fraction of chemically peculiar stars in GCs, which is found to be $\sim70${\percent} \citep{Carretta2010}. One possible solution for this problem is that the cluster loses a significant fraction of its primordial stars, i.e. $\ge95${\percent}, to increase the relative fraction of chemically peculiar stars. Two of the mechanisms which can cause a significant mass-loss in GCs are stellar evolution induced mass-loss (\citealt{D'Ercole2008}; hereafter DE08) and primordial gas expulsion (\citealt{Khalaj2015}; hereafter KB15).

\par It is worthwhile to mention that, the two other mechanisms proposed to explain the origin of abundance anomalies, i.e. binaries \citep{deMink2009} and pollution of pre-main sequence low-mass stars by the ejecta of rapidly evolving primordial stars \citep{Bastian2013} do not suffer from the mass-budget problem if one assumes a close to $100${\percent} binary fraction or a large survival time-scale for pre-main sequence disc of chemically peculiar stars. However, \citet{Bastian2015a} showed all the proposed models overproduce He for a given change in Na and O. Hence none of the scenarios which has been put forth so far is able to explain all the observed GCs.

\par \cite{Bastian2015b} examined an ensemble of 33 Galactic GCs for any correlations between the fraction of chemically peculiar stars and other cluster properties such as the cluster mass and compared them with predictions given in other studies. They argued against the self-enrichment scenario via nucleosynthesis for the Galactic GCs. 

\par \cite{Larsen2012} and \cite{Larsen2014} found a large ratio of GCs to metal poor field stars in dwarf galaxies. Since any mass-loss from GCs will contribute to the overall metallicity distribution of field stars, this discovery imposes an upper limit for the maximum mass that can be lost by a GC in Fornax. This upper limit which is about $4-5$ times the current cluster mass \citep{Larsen2012, deBoer2015} is below the required limit of the significant mass-loss scenario ($\ge95$\percent). In addition, Fornax GCs are very similar to Galactic GCs in terms of abundance anomalies, mass and luminosities \citep{D'Antona2013, Larsen2012, deBoer2015}. This raises the question if significant mass-loss could have happened in Fornax GCs or not.

\par If the speeds by which the unbound stars leave the GC is large enough then these stars can be stripped off by the Milky Way and leave the gravitational field of Fornax, hence do not contribute to the metallicity distribution of field stars. In this case the upper limit of 4-5 times more massive from \cite{Larsen2012} does not apply any more.

\par In this paper we aim to answer this question by doing \Nbody simulations of stars which initially belong to GCs and then become unbound (hereafter runaway stars). We follow their trajectory over a Hubble time to see whether they leave the Fornax dSph or not. The current paper is structured as follows: In the next section we explain the setup and the details of our \Nbody simulations. In \secref{sec:results} we bring the results of the simulations. In \secref{sec:implications} we discuss the implications of our results for the formation scenarios of chemically peculiar stars in dwarf galaxies and conclude our work in \secref{sec:conclusions}.

\section{N-body simulations}\label{sec:simulations}
\subsection{Milky Way and Fornax Models}\label{sec:models}

\begin{table*}
    \centering
    \caption{Adopted models in this study for the density profile of the Fornax dSph given by \equref{eq:rho}. $\gamma_0$, $\gamma_\infty$, $\eta$, $\rs$ and the total mass are directly taken from Table 2 of \citet{Cole2012}. $\Vc$ is the circular velocity derived by numerical integration (\figref{fig:vel}) and tidal radii are derived from \equref{eq:rt0}.}
    \label{tab:params}
    \begin{tabular}{cccccccccc}
        \hline      
        Model  & Total Mass & $\gamma_0$ & $\gamma_\infty$ & $\eta$ & $\rs$ & Tidal Radius & $\Vc \ (500\pc)$  & $\Vc \ (1000\pc)$ & $\Vc \ (3000\pc)$ \\   
        - & $10^8\Msun$ & - & - & - & $\kpc$ &  $\kpc$ & $\kms$  & $\kms$ & $\kms$ \\
        \hline      
        LC & $8.00$ & 0.07 & 4.65 & 3.70 & 1.40 & 6.7 & 13.3 & 24.2 & 30.3\\
        WC & $1.23$ & 0.08 & 4.65 & 2.77 & 0.62 & 3.6 & 15.2 & 17.9 & 12.8 \\
        IC & $1.51$ & 0.13 & 4.24 & 1.37 & 0.55 & 3.8 & 16.0 & 17.3 & 13.6\\
        SC & $1.98$ & 0.52 & 4.27 & 0.93 & 0.80 & 4.2 & 16.0 & 17.2 & 14.6\\
        \hline      
    \end{tabular}   
\end{table*}

\par We use a logarithmic potential field to model the Milky Way halo
\begin{equation}\label{eq:phi}
	\Phi(R) = \frac{1}{2}\Vinfty^2\ln{\left(\Rc^2+R^2\right)}
\end{equation}
where $\Rc$ is the core-radius and $\Vinfty$ is the asymptotic velocity. We assume $\Rc=R_\odot=8300\pc$ and $\Vinfty=V(R_\odot)=239\kms$ \citep{McMillan2011, Irrgang2013}. In a right-handed Galactocentric coordinate system whose $x-$, $y-$ and $z$ axes point towards the Sun, the direction of the orbital motion of the Sun and the North Galactic pole respectively, the current position and velocity vectors of Fornax are $\vec{R}=(-41.3\kpc, -51.0\kpc, -134.1\kpc)$ and $\vec{V}=(-38\kms, -158\kms, 114\kms)$ respectively \citep{Pawlowski2013}. By numerical integration of the Fornax orbit in the Milky Way potential field, given by \equref{eq:phi}, we find that the current distance, perigalactic distance, apogalactic distance, eccentricity, orbital inclination and radial and azimuthal periods of Fornax are $R=149.3\kpc$, $\Rp=99.4\kpc$, $\Ra=154.7\kpc$, $e\equiv(\Ra-\Rp)/(\Ra+\Rp)=0.22$, $i=81\degr$, $T_r=2.4\Gyr$ and $T_\psi=3.3\Gyr$. 

\par For the density profile of Fornax, we use the recent models of \cite{Cole2012} which are derived by applying Markov Chain Monte Carlo (MCMC) analysis to the kinematic data of more than 2000 stars in Fornax. These models have the benefit of including both the stellar and dark matter content of Fornax and that they cover a wide-range of inner density slopes \citep{Cole2012}. The models are designated by large core (LC), weak cusp (WC), intermediate cusp (IC) and steep cusp (SC), all of which have the following form for the density profile
\begin{equation}\label{eq:rho}
	\rho(r)\propto \left(\frac{r}{\rs}\right)^{-\gamma_0}\left(1+\left(\frac{r}{\rs}\right)^{\eta}\right)^{\frac{\gamma_0-\gamma_\infty}{\eta}}{\rm sech}(r/10\kpc)
\end{equation}
where $\rs$ is a scale radius, $\gamma\equiv-{\rm d}\ln{\rho}/{\rm d}r$ is the logarithmic density slope and $\eta$ is a constant. The best-fitting parameters are found using MCMC. \tabref{tab:params} summarizes the characteristics of these models.

\par
\begin{figure}
  \includegraphics[width=0.45\textwidth]{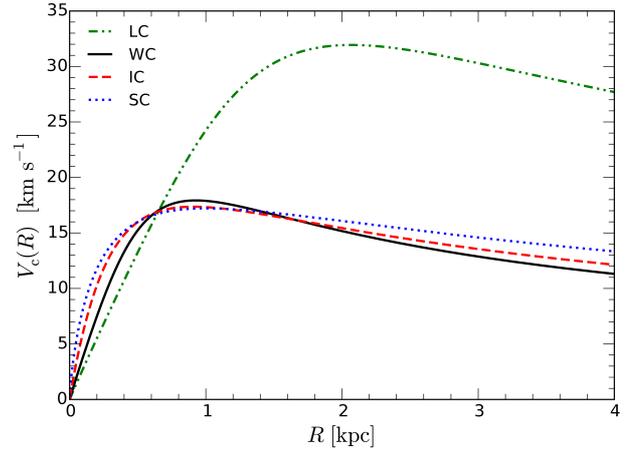}
  \caption{The Fornax velocity curves as a function of radius according to different models given by \equref{eq:rho}. Compared to cuspy models, the large core model predicts a larger circular velocity for $R>0.8\kpc$ due to its larger enclosed mass.}
  \label{fig:vel}
\end{figure}

 The models describe the observed velocity dispersion profile of Fornax very well, however they do not lead to analytical relations for other dynamical parameters such as the enclosed mass at each radius $M(R)$. As a result we used numerical integration to find $M(R)$ (needed for force calculations) and the circular orbital velocity $\Vc(R)$ (needed for initial conditions) and verified our results by comparing with Fig.~2 of \cite{Cole2012}. Figure~\ref{fig:vel} illustrates $\Vc(R)$ for different models adopted in this study. The effect that the choice of the density profile of Fornax has on our results is discussed in \secref{sec:results}.

\subsection{Tidal Radius of Fornax and Escape Criterion}\label{sec:tidal}
For each of the density profiles introduced in \secref{sec:models}, we calculate the tidal radius $\Rt$ of Fornax. The tidal radii are needed later as they specify the ejection criteria from Fornax.

\par The tidal radius of a stellar system with mass $M$, on a circular orbit with radius $\RG$ around a host galaxy with a logarithmic potential specified by \equref{eq:phi}, is
\begin{equation}\label{eq:rt0}
	\Rt = \left(\frac{GM}{2\Vinfty^2}\right)^\frac{1}{3}\RG^\frac{2}{3}\left(\left(\frac{\Rc}{\RG}\right)^2+1\right)^\frac{2}{3}
\end{equation}
The derivation details of this equation are given in Appendix~\ref{sec:A1}.

\par \cite{Baumgardt2003} and \cite{Webb2013, Webb2014} showed that stellar systems with eccentric orbits around a host galaxy have a different tidal radius and mass-loss rate compared to circular orbits. However, by extrapolation of the results presented in \cite{Webb2013}, when the eccentricity is small ($e\leq0.25$), as is the case for Fornax with $e=0.22$, the tidal radius of a stellar system can be approximated by that of a circular orbit with a radius equal to the perigalactic distance of the elliptic orbit. As a result, we approximate the Fornax effective tidal radius by replacing $\RG$ in \equref{eq:rt0} with the perigalactic distance of Fornax orbit ($\Rp=99.4\kpc$). Depending on the model that we have adopted for the density profile of Fornax, we obtain different total masses and different tidal radii which are listed in \tabref{tab:params}. Cuspy models have an average tidal radius of $\Rt\approx3.9\kpc$ (comparable with $\Rt=3.6\kpc$ from \citealt{Walcher2003}), whereas the tidal radius of the LC model is $\Rt=6.7\kpc$ owing to its larger total mass. For comparison, by fitting \citet{King1966} truncated models to the density profile of Fornax, \citet{Irwin1995} derived a tidal radius of $71\farcm1\pm4\farcm0$ for Fornax, which corresponds to a tidal radius of $3.1\pm0.16\kpc$ at the current distance of $149.3\kpc$ (derived in \secref{sec:models}) for Fornax .

\begin{figure}    
  \includegraphics[width=0.45\textwidth]{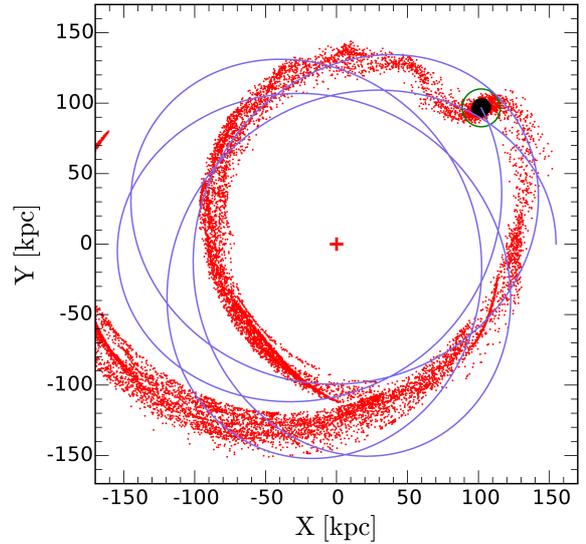} 
  \caption{Spatial distribution of stars in the halo of the Milky Way for the LC model after one Hubble time. The Milky Way is denoted by the central cross at the origin. In this simulation, the initial distance of the stars from the centre of Fornax is $R=3.0\kpc$ and their initial speed is $v=35\kms$. Red points represent stars which have left Fornax (about $70${\percent} mass-loss). The clump of black points enclosed by the green circle shows stars which still reside within the tidal radius of Fornax. The radius of the green circle is equal to 2 times the tidal radius of Fornax at perigalacticon. The Rosette like orbit of Fornax is shown by the blue solid line. In this figure, ${\rm X}$ and ${\rm Y}$ correspond to the coordinate of stars in the orbital plane of Fornax and not the Galactocentric rest frame defined in \secref{sec:models}.}
  \label{fig:stars}
\end{figure}

\begin{figure}
  \includegraphics[width=0.45\textwidth]{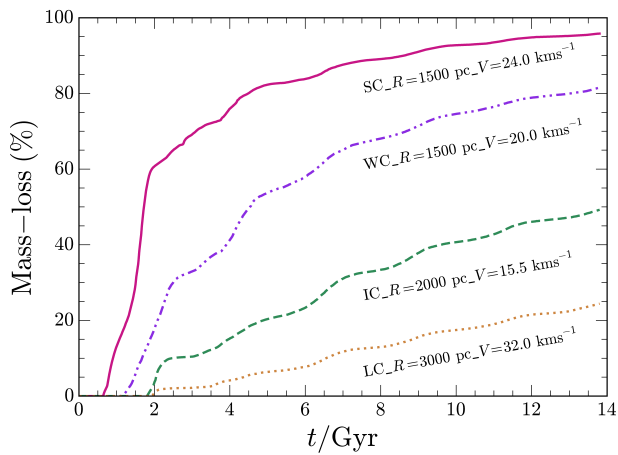}
  \caption{Mass-loss as a function of time for 4 different simulations. The first entry in the label of each simulation is the adopted model for the density profile of Fornax, the second entry is the initial distance of the stars from the centre of Fornax (equal to the orbital radius of GCs) and the last entry is the initial speed of the stars. Note how the LC model has a lower mass-loss compared to other models for $R<3.0\kpc$. For models with significant mass-loss, the major mass-loss happens in the first $5\Gyr$, followed by a slow rise afterwards.}
  \label{fig:evolution}
\end{figure}

\begin{figure}
  \includegraphics[width=0.45\textwidth]{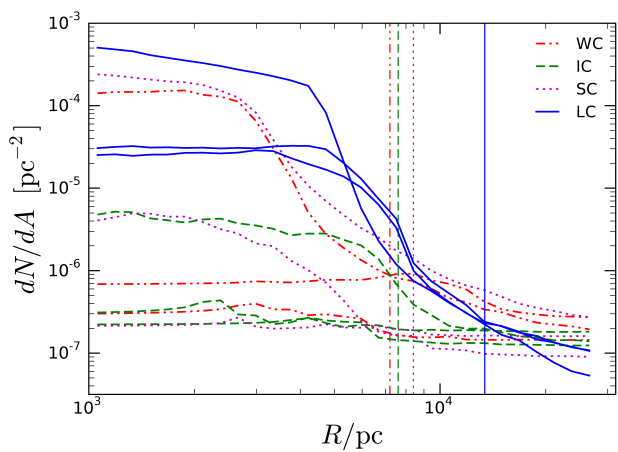}
  \caption{Surface density of stars as a function of projected radius for different models after a Hubble time calculated using infinite projections method \citep{Mashchenko2005}. The vertical lines mark $2\Rt$ where $\Rt$ is the tidal radius of Fornax at perigalacticon according to different models. The densities are almost constant inside Fornax and they drop beyond $\Rt$ and start to level off beyond $2\Rt$ as they reach the background density. Curves with the same color and line style correspond to the same density profile for Fornax but differ in model parameters, i.e. different initial orbital radii and initial speed for stars. The parameters for the models shown in this figure are (from the top to the bottom for each density profile) WC: ($2.5\kpc$, $15\kms$), ($1.5\kpc$, $35\kms$), ($1.0\kpc$, $30\kms$); IC: ($2.5\kpc$, $20\kms$), ($1.5\kpc$, $27\kms$), ($1.0\kpc$, $35\kms$); SC ($1.5\kpc$, $35\kms$), ($2.5\kpc$, $15\kms$), ($1.5\kpc$, $25\kms$); LC: ($4.5\kpc$, $15\kms$), ($5.5\kpc$, $25\kms$), ($3.5\kpc$, $35\kms$).}
  \label{fig:density}
\end{figure}

\subsection{Initial conditions of runaways stars}\label{sec:ic}

\par Our assumption is that runaway stars initially belong to one of the Fornax GCs and become unbound by some unknown mechanism. As a result, they all have the same initial distance from the centre of Fornax. In our simulations, the initial speed of each runaway star corresponds to its excess speed, i.e. its speed when it is infinitely far from the cluster and is no longer effected by its gravitational pull. As a result, we don't consider the gravitational effect of the GCs on these runaway stars. Nor do we consider the gravitational interaction between stars. The runaway stars only feel the gravitational force of Fornax and the Milky Way. 

\par We have $N=20000$ stars in each simulation and give all of them the same initial speed and the same initial distance from the centre of Fornax. The orientation of the velocity vector of each runaway star is determined from its speed and a uniform spherical distribution. For the whole grid of runs we vary the speed of runaway stars from $0.0$ to $50.0\kms$ spaced by $0.5\kms$. The initial distance of the stars varies from $0.5$ to $3.0\kpc$ for cuspy models and $3.0$ to $6.0\kpc$ for the LC model, spaced by $0.5\kpc$. These values are below the tidal radius of Fornax for each density profile. The initial position of Fornax is set to its apogalacticon in each simulation. 

\par We let the simulations run for $T=13.8\Gyr$ and determine the fraction of stars lost from Fornax using the tidal radii found in \secref{sec:tidal}. We consider all the stars with $R>2\Rt$ to find the total mass-loss at each snapshot where $\Rt$ is the tidal radius of Fornax.

\section{Results}\label{sec:results}

\begin{figure*}
  \includegraphics[width=0.9\textwidth]{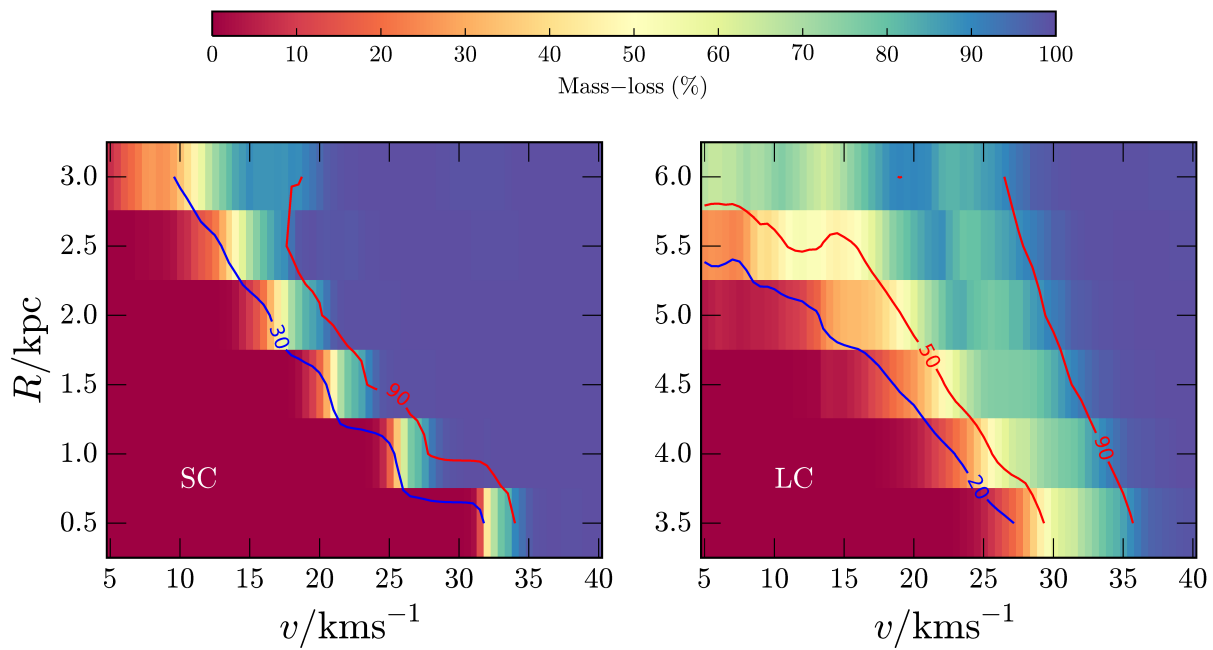}
  \caption{Mass-loss from Fornax as a function of the initial velocity and the initial distance of stars from the centre of Fornax for SC and LC models. WC and IC models have a very similar pattern to the SC model. The LC model predicts a lower mass-loss compared to cuspy models for $R<3.0\kpc$.}
  \label{fig:gridv}
\end{figure*}

\par Figure~\ref{fig:stars} shows the distribution of runaway stars with respect to the Milky Way after a Hubble time for one of the LC simulations, with $\RGC=3.0\kpc$ and $v=35\kms$. According to this figure, a fraction of stars ($\sim30$ {\percent}) will remain in Fornax and the rest leaves Fornax due to their large velocities. The evolution of mass-loss over a Hubble time is depicted in \figref{fig:evolution}. For the runs with a large mass-loss ($>70${\percent}), the majority of runaway stars escape Fornax after only $5\Gyr$ after which the mass-loss increases only slowly.

\par Figure~\ref{fig:density} shows the projected number density of stars calculated using the infinite projections method of \citet{Mashchenko2005} which reduces the random fluctuations caused by low number statistics. The densities are almost constant inside Fornax ($R<\Rt$) and then rapidly decrease outside the tidal radius. Beyond $R>2\Rt$ the density curves tend to a constant value which is equal to the background density.

\par The whole grid of our \Nbody simulations is given in \figref{fig:gridv} which illustrates the mass-loss from the Fornax dSph as a function of initial speed and the initial distance of the stars  from the centre of Fornax. Mass-loss increases for larger radii and larger speeds, which is the direct consequence of Fornax potential field to become weaker at larger radii. Using this grid, one can estimate the mass-loss which could have happened in Fornax, provided that the initial orbital radii of Fornax GCs are known.

\par The current projected distances of Fornax GCs from the centre of the dSph are $1.60\kpc$, $1.05\kpc$, $0.43\kpc$, $0.24\kpc$ and $1.43\kpc$ \citep{Mackey2003} and they are all old, with ages ranging from $7$ to $14\Gyr$ \citep{Strader2003, Mackey2003}. This is puzzling since the GCs are moving in a dark matter halo which can cause dynamical friction and shrink the orbital radius of Fornax GCs within a Hubble time. Yet we don't see any cluster merging in the centre of Fornax and there is no evidence of a bright stellar nucleus in the dSph. This is commonly referred to as the timing problem of Fornax GCs. \citet{Cole2012} studied this problem and generated a wide range of initial conditions for the orbit of Fornax GCs. They showed that, in case of a cuspy density profile for Fornax (i.e. WC, IC and SC models), when the initial orbital radii of Fornax GCs are less than $2-3\kpc$, the orbital radii of Fornax GCs significantly decrease as a result of dynamical friction after only $2\Gyr$. In addition, all clusters except for Fornax 1 will fall into the centre of Fornax over a Hubble time. Since Fornax GCs are located at radii far from the central region of the dSph (i.e. $r>200\pc$), this implies that the initial orbital radii of clusters must have been larger than their current orbital radii, i.e. they all have initial orbital radii of $\ge2\kpc$. Otherwise, they must have fallen into the centre of the dSph and merge by now. This can be seen from figure 4 of \citet{Cole2012} which shows the evolution of the orbital radii of Fornax GCs for different density profiles. In contrast, the infall of the clusters does not happen in case of the LC model. GCs in this model show a dynamical buoyancy instead of dynamical friction as shown by \cite{Cole2012}. It can be inferred from figure~4 of \cite{Cole2012} that Fornax GCs formed more or less at the same radii as they are seen today ($R<1.5\kpc$) in this case.

\par Knowing that the initial orbital radii of GCs must be either $2-3\kpc$ or $\sim1.5\kpc$ depending on the density profile, what remains is to see at what radius the runaway stars are released. The significant mass-loss of the GCs occurs over a time-scale of the order of Myrs (see \secref{sec:implications}). This is much smaller than the dynamical evolution time-scale of the orbits due to dynamical friction/buoyancy which is of the order of Gyrs \citep{Cole2012}. As a result one can assume that when the significant mass-loss happened in Fornax GCs, the runaway stars were released with an initial orbital radius equal to that of the cluster, i.e. $2-3\kpc$ for cuspy models and $\sim1.5\kpc$ for the LC model.

\par Combining these facts with \figref{fig:gridv}, one can see that for all of the cuspy models, mass-loss can be as high as $90${\percent} for $R=2-3\kpc$ and $v>20\kms$, whereas for the large core model this value is less than $10${\percent}. This indicates that the LC model systematically predicts a lower mass-loss at the same radii compared to cuspy models (also evident in \figref{fig:evolution}). To achieve the same mass-loss from Fornax in case of the LC model, the speed of stars need to be $v>35\kms$ at $R=2-3\kpc$. The reason that the LC model has a lower mass-loss is due to its larger enclosed mass at each radius which leads to a larger gravitational force compared to cuspy models. This can also be inferred from \figref{fig:vel} which shows that the LC model has a larger circular velocity for $R>0.8\kpc$. 

\section{Implications for formation scenarios of chemically peculiar stars}\label{sec:implications}
So far, two mechanisms have been suggested to produce a significant mass-loss from GCs: stellar evolution induced mass-loss in extended clusters \citepalias{D'Ercole2008} and primordial gas expulsion \citepalias{Khalaj2015}. In this section, we estimate the mass-loss which can be achieved by both of these mechanisms based on our results given in \secref{sec:results}.

\subsection{Gas expulsion}\label{sec:gas}
\begin{figure}
  \includegraphics[width=0.45\textwidth]{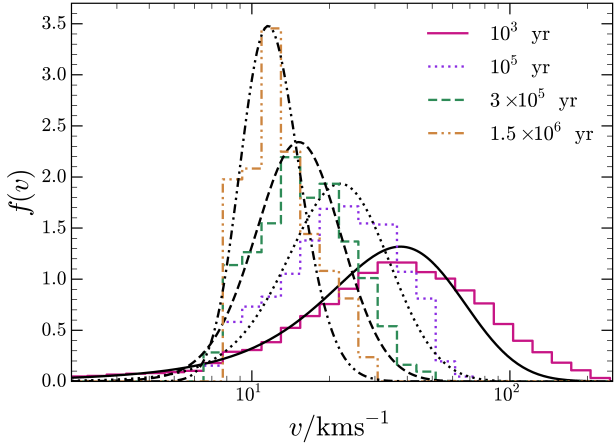}  
  \caption{The normalized distribution of excess speeds of runaway stars for an isolated cluster with an initial stellar mass of $M_\star=10^6\Msun$ and an initial half-mass radius of $\Rh=1.0\pc$ for different gas expulsion time-scales. Data is taken from \citetalias{Khalaj2015}. The black lines show the best Gumbel fit to the data whose probability density function is given by $f(v)=\frac{1}{\sigma}\exp{\left(-\frac{v-\mu}{\sigma}-\exp{\left(-\frac{v-\mu}{\sigma}\right)}\right)}$, where $\mu$ and $\sigma$ are location (mode) and scale parameters of the distribution.}  
  \label{fig:fv}
\end{figure}

\begin{figure*}  
  \includegraphics[width=0.9\textwidth]{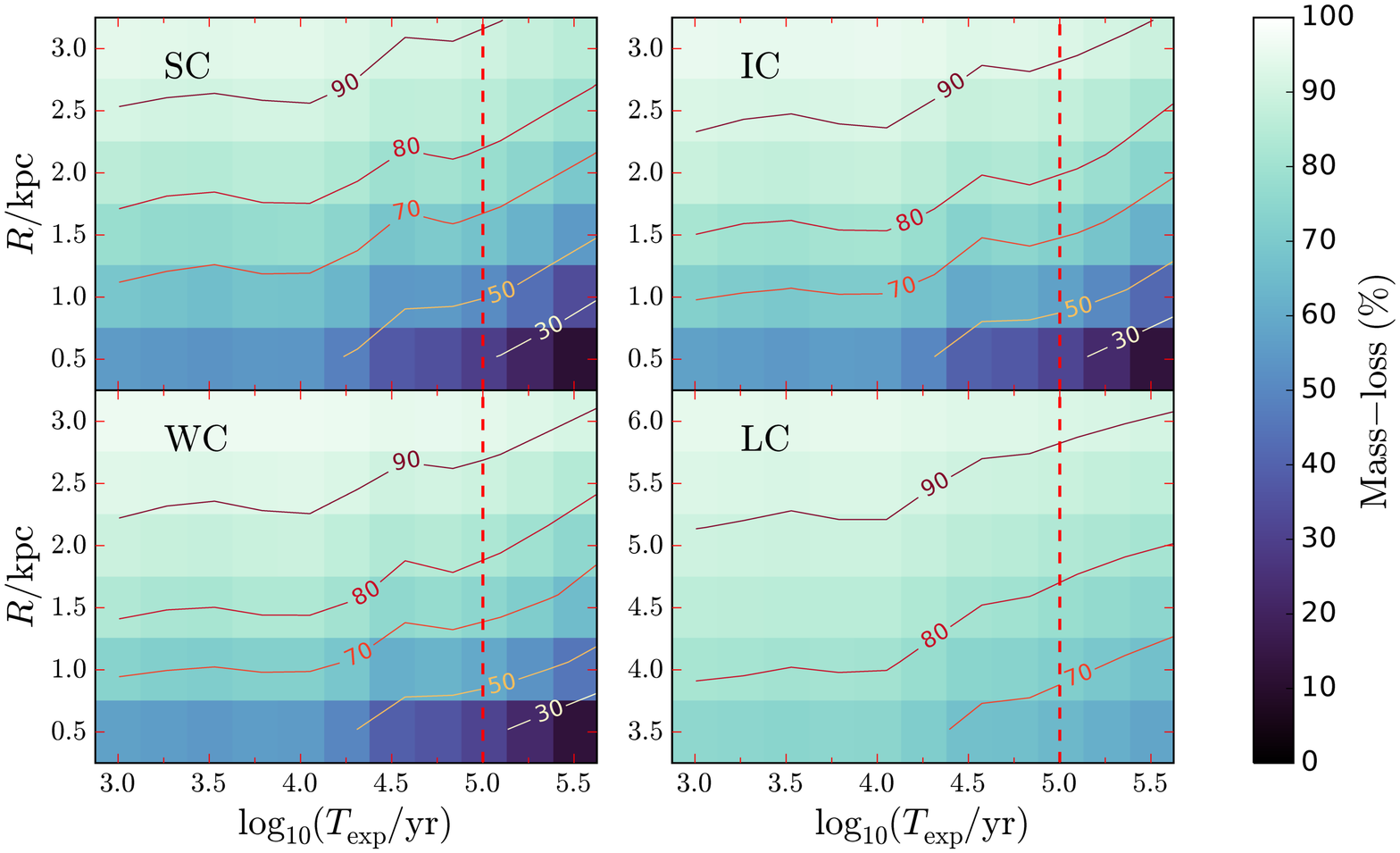}   
  \caption{Mass-loss from Fornax as a function of cluster orbital radius and gas expulsion time-scale for different models. The LC model predicts a lower mass-loss compared to cuspy models for $R<3.0\kpc$. The vertical dashed line marks the upper limit on gas expulsion time-scale found in \citetalias{Khalaj2015}.}
  \label{fig:dist}
\end{figure*}

In this section we test whether we can achieve a high mass-loss from Fornax if runaway stars are a result of gas expulsion in GCs.

\par In \citetalias{Khalaj2015}, we did \Nbody simulations of GCs with two stellar populations to study the role of primordial gas expulsion on the mass-loss. Using a series of Monte Carlo simulations, we then compared the simulated GCs with Galactic clusters and found that gas expulsion induced mass-loss can explain the large fraction of chemically peculiar stars in GCs, as well the observed distribution for the current mass and the half-mass radius of the Galactic GCs. Assuming that all Galactic GCs have undergone gas expulsion induced mass-loss, we found an upper limit of $10^5\yr$ for the gas expulsion time scale. In addition, we found that the typical initial stellar masses, half-mass radii and the initial mass of the primordial gas left in the Galactic GCs need to be around $10^6\Msun$, $1.0\pc$ and $10^6\Msun$ respectively, so the total mass of each cluster is $\sim2\times10^6\Msun$.

\par To estimate the mass-loss from Fornax as a result of gas expulsion in Fornax GCs, we use the \citetalias{Khalaj2015} grid of \Nbody simulations. Our assumption is that the initial mass and half mass radii of Fornax GCs are similar to those of Galactic GCs, i.e. $\Rh=1.0\pc$ and $\Mc=2\times10^6\Msun$. We then need the ratio of tidal radii of the GCs to their half-mass radii to completely identify the location of Fornax GCs in the \Nbody grid of \citetalias{Khalaj2015}. We calculate the tidal radius of the GCs in the potential field of Fornax using the following equation

\begin{equation}\label{eq:rt_num0}
\Rt = \Mc^{\frac{1}{3}}\left(3\frac{M(\RG)}{\RG^3}-4\pi\rho(\RG)\right)^{-\frac{1}{3}} 
\end{equation}
where $\Mc=2\times10^6\Msun$ is the total mass of each cluster and $M(\RG)$ is the enclosed mass of Fornax within $\RG$ and needs to be found by numerical integration. The derivation details of this equation are given in Appendix~\ref{sec:A2}.

\par Using \equref{eq:rt_num0}, we found that the tidal radii of our clusters in the tidal field of Fornax are larger than $100\pc$ for orbital radii $R>0.5\kpc$ for all density profiles of Fornax. Such a large tidal radius corresponds to an almost isolated cluster as can be seen from figures~1 to 3 of \citetalias{Khalaj2015}. As a result we take the data of isolated clusters from \citetalias{Khalaj2015} and find the distribution of excess speeds for runaway stars. Figure~\ref{fig:fv} shows the normalized distribution of excess speeds for different gas expulsion time-scales. As it can been seen from this figure, lower gas expulsion time-scales lead to faster excess speeds. In addition, for $\Texp\le10^5\yr$, they have a mean excess speed larger than $20\kms$, indicating that it is possible to achieve a significant mass-loss with gas expulsion.

\par To determine the extent of mass-loss as a function of the gas expulsion time-scale and the orbital radii of GCs, we did an analysis as follows: For each runaway star we randomly draw a value from the excess speed distribution assuming a particular gas expulsion time-scale (\figref{fig:fv}). We then find the velocity vector of the runaway star using its excess speed and a uniform spherical distribution for its orientation in space. We then add the velocity of the star to the velocity of the GC around Fornax at each radius (circular velocity) and the velocity of Fornax around the Milky Way. Finally, we derive the speed distribution of the runaway stars from their velocity vectors and convolve this distribution with our grid in \figref{fig:gridv}.

\par Figure~\ref{fig:dist} shows the mass-loss as a function of the GCs orbital radii and gas expulsion time-scales. As it can be seen from this figure, depending on the density profile of Fornax, the gas expulsion time-scale and the initial orbital radius of the cluster, mass-loss from Fornax varies from $30$ to more than $90${\percent}. As a result, with gas expulsion, it is possible that the majority of the runaway stars leave the Fornax galaxy without contributing to the metallicity distribution of field stars. The implication of these results is further discussed in section \ref{sec:conclusions}.

\subsection{Stellar evolution induced mass-loss}
We compare the results obtained in \secref{sec:gas} with the \citetalias{D'Ercole2008} models for clusters with $\Mc=10^6\Msun$ to find the typical excess speed of runaway stars which become unbound as a result of stellar evolution. \citetalias{D'Ercole2008} used hydrodynamical and \Nbody simulations, to model the formation and dynamical evolution of multiple stellar populations in GCs, based on the AGB scenario and the stellar evolution induced mass-loss caused by the supernovae ejecta of primordial stars. In their models, GCs have \citet{King1966} density profiles with core, half-mass and tidal radii of $\Rc=2.85$, $\Rh\sim10.3\pc$ and $\Rt=90\pc$ respectively and they lose a substantial amount of their mass during a period of $10-100$ dynamical times. These models can explain the large fraction of chemically peculiar stars in GCs. However, one needs very extended clusters ($\Rh\sim10.3\pc$) to lose a substantial amount of mass.

\par To estimate the typical excess speed of runaway stars in \citetalias{D'Ercole2008} models, we take the typical excess speeds found in \secref{sec:gas} and then scale them to the models of \citetalias{D'Ercole2008}, using the mass and the radius of the clusters. The dynamical time-scale of a cluster and the velocity dispersion of its stars is proportional to $\Tdyn\propto M^{-1/2}R^{3/2}$ and $\sigma_v\propto (M/R)^{1/2}$ respectively. The models that we studied in \secref{sec:gas} have $\Mc=2\times10^6\Msun$, $\Rh=1.0\pc$ and $\Tdyn=3\times10^4\yr$. This corresponds to a mass and radius ratio of $\Mc({\rm DE08})/\Mc({\rm KB15}) = (10^6\Msun)/(2\times10^6\Msun) = 0.5$ and $\Rh({\rm DE08})/\Rh({\rm KB15}) = (10.3\pc)/(1.0\pc) = 10.3$ respectively. As a result, the velocity dispersion of stars in \citetalias{D'Ercole2008} models is smaller than that of \citetalias{Khalaj2015} by a factor of $(0.5/10.3)^{-1/2}\sim4.5$, and the dynamical time-scale of the clusters is larger by a factor of $0.5^{-1/2}\times10.3^{3/2}\sim47$. In case of impulsive mass-loss, these values correspond to a mass-loss time-scale of at least $47\times3\times10^4\yr=1.5\Myr$. In \figref{fig:fv}, the dash-dotted curve corresponds to a gas expulsion time-scale of $T=1.5\Myr$. According to this curve the typical excess speed of stars is $\sim12\kms$ for \citet{Khalaj2015} models. We found that the velocity dispersion of stars in the models of \citetalias{D'Ercole2008} is smaller by a factor of $\sim4.5$, this translates into a typical excess speed of $12/4.5=2.6\kms$ for \citetalias{D'Ercole2008} models. Assuming that the unbound stars have a speed distribution similar to \figref{fig:fv} and applying the same procedure explained in \secref{sec:gas}, we obtain a mass-loss of less than $\sim30${\percent} from Fornax for $R=3.0\kpc$ and cuspy models. This upper limit on mass-loss which is smaller for clusters with $R<3.0\kpc$, is significantly lower than what we obtained for gas expulsion, indicating that the mechanism proposed by \citetalias{D'Ercole2008} is not efficient in expelling unbound stars from the Fornax dSph.

\section{Discussion and Conclusions}\label{sec:conclusions}
Using a series of N-body simulations we determined the mass-loss from the Fornax dSph as a function of the initial speed and the initial distance of stars from the Fornax centre. We assumed a set of cuspy and cored models and tracked the trajectory of stars over a Hubble time. We then combined our results with those of \citet{D'Ercole2008} and \citet{Khalaj2015} to test whether significant mass-loss could have happened in Fornax GCs. Our results show that gas expulsion induced mass-loss in contrast to stellar evolution induced mass-loss can lead to large excess speeds for runaway stars and larger mass-loss from Fornax.

\par In particular, in case of gas expulsion induced mass-loss in GCs and for cuspy models when the orbital radii of GCs are between $2-3\kpc$, a mass-loss of $\sim90${\percent} can be achieved from Fornax. Comparing this result with the fact that Fornax GCs could have been 4-5 times more massive initially \citep{Larsen2012}, means that Fornax GCs could have lost a significant fraction ($>95$\percent) of their initial mass and still yield a metallicity distribution of field stars which is consistent with observations. This result is especially important for Fornax 3 which is the most massive cluster in Fornax and imposes the tightest limit on the mass-loss from GCs in this galaxy. 

\par In contrast, this is not the case for the LC model. GCs in this model, as shown by \citet{Cole2012} formed at the same radii as they are seen today ($R<1.5\kpc$) in which case the maximum mass-loss attained for LC model will be less than $50$\percent. A large core with $\Rc\sim1.5\kpc$, overestimates the overall mass of the Fornax which is found to be $\sim1-3\times10^8\Msun$ by several studies (e.g. \citealt{Irwin1995, Walcher2003, Lokas2009}). In addition, \cite{Strigari2006} showed that such a large core for Fornax will be inconsistent with standard warm and dark matter models and predicts an unreasonably massive dark matter halo and a large asymptotic velocity of $200\kms$. However there are studies which prefer a cored model of $\Rc\sim1.0\kpc$ (e.g. \citealt{Amorisco2013}) and it seems that different cored-cuspy models are possible to describe the dark matter profile of Fornax.

\par We showed that, if Fornax has a cuspy density profile or a small core, then it is possible to reconcile the predictions of significant mass-loss scenario \citep{Khalaj2015} with the observed metallicity distribution of field stars in Fornax \citep{Larsen2012}.

\par We have done our simulations based on the assumption that the potential field of the Milky Way and Fornax are constant in time. This might not be true, especially for Fornax which has been suggested to be the remnant of a recent merging event \citep{Coleman2004, Yozin2012, delPino2015}. In such a case, the present-day metallicity distribution of field stars does not directly reflect the mass-loss history of GCs in  Fornax and the upper limit for the initial mass of Fornax GCs estimated by \citet{Larsen2012} does not apply.


\section*{Acknowledgements}
We would like to thank the anonymous referee for his/her useful comments and suggestions which improved the quality of this work. PK and HB are members of international team 271 "Massive star clusters across the Hubble time" led by Corinne Charbonnel in the International Space Science Institute (ISSI), Bern, Switzerland. PK and HB would like to appreciate ISSI funding and hospitality during the team meetings in January 2014 and June 2015.


\bibliographystyle{mnras}

\begin{thebibliography}{}
\makeatletter
\relax
\def\mn@urlcharsother{\let\do\@makeother \do\$\do\&\do\#\do\^\do\_\do\%\do\~}
\def\mn@doi{\begingroup\mn@urlcharsother \@ifnextchar [ {\mn@doi@}
  {\mn@doi@[]}}
\def\mn@doi@[#1]#2{\def\@tempa{#1}\ifx\@tempa\@empty \href
  {http://dx.doi.org/#2} {doi:#2}\else \href {http://dx.doi.org/#2} {#1}\fi
  \endgroup}
\def\mn@eprint#1#2{\mn@eprint@#1:#2::\@nil}
\def\mn@eprint@arXiv#1{\href {http://arxiv.org/abs/#1} {{\tt arXiv:#1}}}
\def\mn@eprint@dblp#1{\href {http://dblp.uni-trier.de/rec/bibtex/#1.xml}
  {dblp:#1}}
\def\mn@eprint@#1:#2:#3:#4\@nil{\def\@tempa {#1}\def\@tempb {#2}\def\@tempc
  {#3}\ifx \@tempc \@empty \let \@tempc \@tempb \let \@tempb \@tempa \fi \ifx
  \@tempb \@empty \def\@tempb {arXiv}\fi \@ifundefined
  {mn@eprint@\@tempb}{\@tempb:\@tempc}{\expandafter \expandafter \csname
  mn@eprint@\@tempb\endcsname \expandafter{\@tempc}}}

\bibitem[\protect\citeauthoryear{{Amorisco}, {Agnello}  \& {Evans}}{{Amorisco}
  et~al.}{2013}]{Amorisco2013}
{Amorisco} N.~C.,  {Agnello} A.,   {Evans} N.~W.,  2013, \mn@doi [\mnras]
  {10.1093/mnrasl/sls031}, \href
  {http://adsabs.harvard.edu/abs/2013MNRAS.429L..89A} {429, L89}

\bibitem[\protect\citeauthoryear{{Bastian} \& {Lardo}}{{Bastian} \&
  {Lardo}}{2015}]{Bastian2015b}
{Bastian} N.,  {Lardo} C.,  2015, \mn@doi [\mnras] {10.1093/mnras/stv1661},
  \href {http://adsabs.harvard.edu/abs/2015MNRAS.453..357B} {453, 357}

\bibitem[\protect\citeauthoryear{{Bastian}, {Lamers}, {de Mink}, {Longmore},
  {Goodwin}  \& {Gieles}}{{Bastian} et~al.}{2013}]{Bastian2013}
{Bastian} N.,  {Lamers} H.~J.~G.~L.~M.,  {de Mink} S.~E.,  {Longmore} S.~N.,
  {Goodwin} S.~P.,   {Gieles} M.,  2013, \mn@doi [\mnras]
  {10.1093/mnras/stt1745}, \href
  {http://adsabs.harvard.edu/abs/2013MNRAS.436.2398B} {436, 2398}

\bibitem[\protect\citeauthoryear{{Bastian}, {Cabrera-Ziri}  \&
  {Salaris}}{{Bastian} et~al.}{2015}]{Bastian2015a}
{Bastian} N.,  {Cabrera-Ziri} I.,   {Salaris} M.,  2015, \mn@doi [\mnras]
  {10.1093/mnras/stv543}, \href
  {http://adsabs.harvard.edu/abs/2015MNRAS.449.3333B} {449, 3333}

\bibitem[\protect\citeauthoryear{{Baumgardt} \& {Makino}}{{Baumgardt} \&
  {Makino}}{2003}]{Baumgardt2003}
{Baumgardt} H.,  {Makino} J.,  2003, \mn@doi [\mnras]
  {10.1046/j.1365-8711.2003.06286.x}, \href
  {http://adsabs.harvard.edu/abs/2003MNRAS.340..227B} {340, 227}

\bibitem[\protect\citeauthoryear{{Bertin} \& {Varri}}{{Bertin} \&
  {Varri}}{2008}]{Bertin2008}
{Bertin} G.,  {Varri} A.~L.,  2008, \mn@doi [\apj] {10.1086/592684}, \href
  {http://adsabs.harvard.edu/abs/2008ApJ...689.1005B} {689, 1005}

\bibitem[\protect\citeauthoryear{{Carretta}, {Bragaglia}, {Gratton},
  {Recio-Blanco}, {Lucatello}, {D'Orazi}  \& {Cassisi}}{{Carretta}
  et~al.}{2010}]{Carretta2010}
{Carretta} E.,  {Bragaglia} A.,  {Gratton} R.~G.,  {Recio-Blanco} A.,
  {Lucatello} S.,  {D'Orazi} V.,   {Cassisi} S.,  2010, \mn@doi [\aap]
  {10.1051/0004-6361/200913451}, \href
  {http://adsabs.harvard.edu/abs/2010A%26A...516A..55C} {516, A55}

\bibitem[\protect\citeauthoryear{{Carretta} et~al.,}{{Carretta}
  et~al.}{2015}]{Carretta2015}
{Carretta} E.,  et~al., 2015, \mn@doi [\aap] {10.1051/0004-6361/201525951},
  \href {http://adsabs.harvard.edu/abs/2015A%26A...578A.116C} {578, A116}

\bibitem[\protect\citeauthoryear{{Cole}, {Dehnen}, {Read}  \&
  {Wilkinson}}{{Cole} et~al.}{2012}]{Cole2012}
{Cole} D.~R.,  {Dehnen} W.,  {Read} J.~I.,   {Wilkinson} M.~I.,  2012, \mn@doi
  [\mnras] {10.1111/j.1365-2966.2012.21885.x}, \href
  {http://adsabs.harvard.edu/abs/2012MNRAS.426..601C} {426, 601}

\bibitem[\protect\citeauthoryear{{Coleman}, {Da Costa}, {Bland-Hawthorn},
  {Mart{\'{\i}}nez-Delgado}, {Freeman}  \& {Malin}}{{Coleman}
  et~al.}{2004}]{Coleman2004}
{Coleman} M.,  {Da Costa} G.~S.,  {Bland-Hawthorn} J.,
  {Mart{\'{\i}}nez-Delgado} D.,  {Freeman} K.~C.,   {Malin} D.,  2004, \mn@doi
  [\aj] {10.1086/381298}, \href
  {http://adsabs.harvard.edu/abs/2004AJ....127..832C} {127, 832}

\bibitem[\protect\citeauthoryear{{D'Antona}, {Caloi}, {D'Ercole}, {Tailo},
  {Vesperini}, {Ventura}  \& {Di Criscienzo}}{{D'Antona}
  et~al.}{2013}]{D'Antona2013}
{D'Antona} F.,  {Caloi} V.,  {D'Ercole} A.,  {Tailo} M.,  {Vesperini} E.,
  {Ventura} P.,   {Di Criscienzo} M.,  2013, \mn@doi [\mnras]
  {10.1093/mnras/stt1057}, \href
  {http://adsabs.harvard.edu/abs/2013MNRAS.434.1138D} {434, 1138}

\bibitem[\protect\citeauthoryear{{D'Ercole}, {Vesperini}, {D'Antona},
  {McMillan}  \& {Recchi}}{{D'Ercole} et~al.}{2008}]{D'Ercole2008}
{D'Ercole} A.,  {Vesperini} E.,  {D'Antona} F.,  {McMillan} S.~L.~W.,
  {Recchi} S.,  2008, \mn@doi [\mnras] {10.1111/j.1365-2966.2008.13915.x},
  \href {http://adsabs.harvard.edu/abs/2008MNRAS.391..825D} {391, 825}

\bibitem[\protect\citeauthoryear{{Decressin}, {Meynet}, {Charbonnel},
  {Prantzos}  \& {Ekstr{\"o}m}}{{Decressin} et~al.}{2007a}]{Decressin2007a}
{Decressin} T.,  {Meynet} G.,  {Charbonnel} C.,  {Prantzos} N.,   {Ekstr{\"o}m}
  S.,  2007a, \mn@doi [\aap] {10.1051/0004-6361:20066013}, \href
  {http://adsabs.harvard.edu/abs/2007A%26A...464.1029D} {464, 1029}

\bibitem[\protect\citeauthoryear{{Decressin}, {Charbonnel}  \&
  {Meynet}}{{Decressin} et~al.}{2007b}]{Decressin2007b}
{Decressin} T.,  {Charbonnel} C.,   {Meynet} G.,  2007b, \mn@doi [\aap]
  {10.1051/0004-6361:20078425}, \href
  {http://adsabs.harvard.edu/abs/2007A%26A...475..859D} {475, 859}

\bibitem[\protect\citeauthoryear{{Irrgang}, {Wilcox}, {Tucker}  \&
  {Schiefelbein}}{{Irrgang} et~al.}{2013}]{Irrgang2013}
{Irrgang} A.,  {Wilcox} B.,  {Tucker} E.,   {Schiefelbein} L.,  2013, \mn@doi
  [\aap] {10.1051/0004-6361/201220540}, \href
  {http://adsabs.harvard.edu/abs/2013A%26A...549A.137I} {549, A137}

\bibitem[\protect\citeauthoryear{{Irwin} \& {Hatzidimitriou}}{{Irwin} \&
  {Hatzidimitriou}}{1995}]{Irwin1995}
{Irwin} M.,  {Hatzidimitriou} D.,  1995, \mnras, \href
  {http://adsabs.harvard.edu/abs/1995MNRAS.277.1354I} {277, 1354}

\bibitem[\protect\citeauthoryear{{Khalaj} \& {Baumgardt}}{{Khalaj} \&
  {Baumgardt}}{2015}]{Khalaj2015}
{Khalaj} P.,  {Baumgardt} H.,  2015, \mn@doi [\mnras] {10.1093/mnras/stv1356},
  \href {http://adsabs.harvard.edu/abs/2015MNRAS.452..924K} {452, 924}

\bibitem[\protect\citeauthoryear{{King}}{{King}}{1966}]{King1966}
{King} I.~R.,  1966, \mn@doi [\aj] {10.1086/109857}, \href
  {http://adsabs.harvard.edu/abs/1966AJ.....71...64K} {71, 64}

\bibitem[\protect\citeauthoryear{{Krause}, {Charbonnel}, {Decressin}, {Meynet}
  \& {Prantzos}}{{Krause} et~al.}{2013}]{Krause2013}
{Krause} M.,  {Charbonnel} C.,  {Decressin} T.,  {Meynet} G.,   {Prantzos} N.,
  2013, \mn@doi [\aap] {10.1051/0004-6361/201220694}, \href
  {http://adsabs.harvard.edu/abs/2013A%26A...552A.121K} {552, A121}

\bibitem[\protect\citeauthoryear{{Kroupa}}{{Kroupa}}{2001}]{Kroupa2001}
{Kroupa} P.,  2001, \mn@doi [\mnras] {10.1046/j.1365-8711.2001.04022.x}, \href
  {http://adsabs.harvard.edu/abs/2001MNRAS.322..231K} {322, 231}

\bibitem[\protect\citeauthoryear{{Larsen}, {Strader}  \& {Brodie}}{{Larsen}
  et~al.}{2012}]{Larsen2012}
{Larsen} S.~S.,  {Strader} J.,   {Brodie} J.~P.,  2012, \mn@doi [\aap]
  {10.1051/0004-6361/201219897}, \href
  {http://adsabs.harvard.edu/abs/2012A%26A...544L..14L} {544, L14}

\bibitem[\protect\citeauthoryear{{Larsen}, {Brodie}, {Forbes}  \&
  {Strader}}{{Larsen} et~al.}{2014}]{Larsen2014}
{Larsen} S.~S.,  {Brodie} J.~P.,  {Forbes} D.~A.,   {Strader} J.,  2014,
  \mn@doi [\aap] {10.1051/0004-6361/201322672}, \href
  {http://adsabs.harvard.edu/abs/2014A%26A...565A..98L} {565, A98}

\bibitem[\protect\citeauthoryear{{{\L}okas}}{{{\L}okas}}{2009}]{Lokas2009}
{{\L}okas} E.~L.,  2009, \mn@doi [\mnras] {10.1111/j.1745-3933.2009.00620.x},
  \href {http://adsabs.harvard.edu/abs/2009MNRAS.394L.102L} {394, L102}

\bibitem[\protect\citeauthoryear{{Mackey} \& {Gilmore}}{{Mackey} \&
  {Gilmore}}{2003}]{Mackey2003}
{Mackey} A.~D.,  {Gilmore} G.~F.,  2003, \mn@doi [\mnras]
  {10.1046/j.1365-8711.2003.06275.x}, \href
  {http://adsabs.harvard.edu/abs/2003MNRAS.340..175M} {340, 175}

\bibitem[\protect\citeauthoryear{{Mashchenko} \& {Sills}}{{Mashchenko} \&
  {Sills}}{2005}]{Mashchenko2005}
{Mashchenko} S.,  {Sills} A.,  2005, \mn@doi [\apj] {10.1086/426132}, \href
  {http://adsabs.harvard.edu/abs/2005ApJ...619..243M} {619, 243}

\bibitem[\protect\citeauthoryear{{McMillan}}{{McMillan}}{2011}]{McMillan2011}
{McMillan} P.~J.,  2011, \mn@doi [\mnras] {10.1111/j.1365-2966.2011.18564.x},
  \href {http://adsabs.harvard.edu/abs/2011MNRAS.414.2446M} {414, 2446}

\bibitem[\protect\citeauthoryear{{Milone} et~al.,}{{Milone}
  et~al.}{2015}]{Milone2015}
{Milone} A.~P.,  et~al., 2015, \mn@doi [\apj] {10.1088/0004-637X/808/1/51},
  \href {http://adsabs.harvard.edu/abs/2015ApJ...808...51Mn} {808, 51}

\bibitem[\protect\citeauthoryear{{Pawlowski} \& {Kroupa}}{{Pawlowski} \&
  {Kroupa}}{2013}]{Pawlowski2013}
{Pawlowski} M.~S.,  {Kroupa} P.,  2013, \mn@doi [\mnras]
  {10.1093/mnras/stt1429}, \href
  {http://adsabs.harvard.edu/abs/2013MNRAS.435.2116P} {435, 2116}

\bibitem[\protect\citeauthoryear{{Piotto} et~al.,}{{Piotto}
  et~al.}{2007}]{Piotto2007}
{Piotto} G.,  et~al., 2007, \mn@doi [\apjl] {10.1086/518503}, \href
  {http://adsabs.harvard.edu/abs/2007ApJ...661L..53P} {661, L53}

\bibitem[\protect\citeauthoryear{{Prantzos} \& {Charbonnel}}{{Prantzos} \&
  {Charbonnel}}{2006}]{Prantzos2006}
{Prantzos} N.,  {Charbonnel} C.,  2006, \mn@doi [\aap]
  {10.1051/0004-6361:20065374}, \href
  {http://adsabs.harvard.edu/abs/2006A%26A...458..135P} {458, 135}

\bibitem[\protect\citeauthoryear{{Renaud}, {Gieles}  \& {Boily}}{{Renaud}
  et~al.}{2011}]{Renaud2011}
{Renaud} F.,  {Gieles} M.,   {Boily} C.~M.,  2011, \mn@doi [\mnras]
  {10.1111/j.1365-2966.2011.19531.x}, \href
  {http://adsabs.harvard.edu/abs/2011MNRAS.418..759R} {418, 759}

\bibitem[\protect\citeauthoryear{{Rey}, {Lee}, {Ree}, {Joo}, {Sohn}  \&
  {Walker}}{{Rey} et~al.}{2004}]{Rey2004}
{Rey} S.-C.,  {Lee} Y.-W.,  {Ree} C.~H.,  {Joo} J.-M.,  {Sohn} Y.-J.,
  {Walker} A.~R.,  2004, \mn@doi [\aj] {10.1086/380942}, \href
  {http://adsabs.harvard.edu/abs/2004AJ....127..958R} {127, 958}

\bibitem[\protect\citeauthoryear{{Strader}, {Brodie}, {Forbes}, {Beasley}  \&
  {Huchra}}{{Strader} et~al.}{2003}]{Strader2003}
{Strader} J.,  {Brodie} J.~P.,  {Forbes} D.~A.,  {Beasley} M.~A.,   {Huchra}
  J.~P.,  2003, \mn@doi [\aj] {10.1086/367599}, \href
  {http://adsabs.harvard.edu/abs/2003AJ....125.1291S} {125, 1291}

\bibitem[\protect\citeauthoryear{{Strigari}, {Bullock}, {Kaplinghat},
  {Kravtsov}, {Gnedin}, {Abazajian}  \& {Klypin}}{{Strigari}
  et~al.}{2006}]{Strigari2006}
{Strigari} L.~E.,  {Bullock} J.~S.,  {Kaplinghat} M.,  {Kravtsov} A.~V.,
  {Gnedin} O.~Y.,  {Abazajian} K.,   {Klypin} A.~A.,  2006, \mn@doi [\apj]
  {10.1086/506381}, \href {http://adsabs.harvard.edu/abs/2006ApJ...652..306S}
  {652, 306}

\bibitem[\protect\citeauthoryear{{Ventura}, {D'Antona}, {Mazzitelli}  \&
  {Gratton}}{{Ventura} et~al.}{2001}]{Ventura2001}
{Ventura} P.,  {D'Antona} F.,  {Mazzitelli} I.,   {Gratton} R.,  2001, \mn@doi
  [\apjl] {10.1086/319496}, \href
  {http://adsabs.harvard.edu/abs/2001ApJ...550L..65V} {550, L65}

\bibitem[\protect\citeauthoryear{{Walcher}, {Fried}, {Burkert}  \&
  {Klessen}}{{Walcher} et~al.}{2003}]{Walcher2003}
{Walcher} C.~J.,  {Fried} J.~W.,  {Burkert} A.,   {Klessen} R.~S.,  2003,
  \mn@doi [\aap] {10.1051/0004-6361:20030768}, \href
  {http://adsabs.harvard.edu/abs/2003A%26A...406..847W} {406, 847}

\bibitem[\protect\citeauthoryear{{Webb}, {Harris}, {Sills}  \& {Hurley}}{{Webb}
  et~al.}{2013}]{Webb2013}
{Webb} J.~J.,  {Harris} W.~E.,  {Sills} A.,   {Hurley} J.~R.,  2013, \mn@doi
  [\apj] {10.1088/0004-637X/764/2/124}, \href
  {http://adsabs.harvard.edu/abs/2013ApJ...764..124W} {764, 124}

\bibitem[\protect\citeauthoryear{{Webb}, {Leigh}, {Sills}, {Harris}  \&
  {Hurley}}{{Webb} et~al.}{2014}]{Webb2014}
{Webb} J.~J.,  {Leigh} N.,  {Sills} A.,  {Harris} W.~E.,   {Hurley} J.~R.,
  2014, \mn@doi [\mnras] {10.1093/mnras/stu961}, \href
  {http://adsabs.harvard.edu/abs/2014MNRAS.442.1569W} {442, 1569}

\bibitem[\protect\citeauthoryear{{Yozin} \& {Bekki}}{{Yozin} \&
  {Bekki}}{2012}]{Yozin2012}
{Yozin} C.,  {Bekki} K.,  2012, \mn@doi [\apjl] {10.1088/2041-8205/756/1/L18},
  \href {http://adsabs.harvard.edu/abs/2012ApJ...756L..18Y} {756, L18}

\bibitem[\protect\citeauthoryear{{de Boer} \& {Fraser}}{{de Boer} \&
  {Fraser}}{2015}]{deBoer2015}
{de Boer} T.~J.~L.,  {Fraser} M.,  2015, preprint, \href
  {http://adsabs.harvard.edu/abs/2015arXiv151005642D} {} (\mn@eprint {arXiv}
  {1510.05642})

\bibitem[\protect\citeauthoryear{{de Mink}, {Pols}, {Langer}  \& {Izzard}}{{de
  Mink} et~al.}{2009}]{deMink2009}
{de Mink} S.~E.,  {Pols} O.~R.,  {Langer} N.,   {Izzard} R.~G.,  2009, \mn@doi
  [\aap] {10.1051/0004-6361/200913205}, \href
  {http://adsabs.harvard.edu/abs/2009A%26A...507L...1D} {507, L1}

\bibitem[\protect\citeauthoryear{{del Pino}, {Aparicio}  \& {Hidalgo}}{{del
  Pino} et~al.}{2015}]{delPino2015}
{del Pino} A.,  {Aparicio} A.,   {Hidalgo} S.~L.,  2015, \mn@doi [\mnras]
  {10.1093/mnras/stv2174}, \href
  {http://adsabs.harvard.edu/abs/2015MNRAS.454.3996D} {454, 3996}

\makeatother
\end{thebibliography}
\input{manuscript.bbl}

\appendix

\section{Tidal Radius of Fornax}\label{sec:A1}
\par Following the approach and the notation of \cite{Renaud2011}, the tidal radius {\footnote{The distance of the first Lagrangian point ($L_1$) from the center of the system.} of a stellar system with mass $M$, on a circular orbit with radius $\RG$ around a host galaxy with an arbitrary potential field $\phiG$, can be written as follows (equations 10 and 17 therein)
\begin{equation}\label{eq:rt_lambda}
\Rt = \left(\frac{GM}{\lambda_{e,1}}\right)^{\frac{1}{3}}
\end{equation}
where $\lambda_{e,1}$ is the largest eigenvalue of the effective tidal tensor{\footnote{The sum of the tidal and centrifugal potentials.}} and is given by
\begin{equation}\label{eq:lambda}
	\lambda_{e,1} = \left(-\frac{\partial^2\phiG}{\partial x^2}\right)_{\RG} - \left(-\frac{\partial^2\phiG}{\partial z^2}\right)_{\RG}
\end{equation}
\par In the derivation of this equation, it has been assumed that the potential of the stellar system is simply $-G\Mc/r$ which is a reasonable assumption at the tidal radius \citep{Renaud2011}.

\par The substitution of the equation \eqref{eq:phi} into equation \eqref{eq:lambda} yields
\begin{equation}
	\lambda_{e,1} = \frac{2\RG^2\Vinfty^2}{\left(\Rc^2+\RG^2\right)^2}
\end{equation}
hence the tidal radius is
\begin{equation}\label{eq:rt}
	\Rt = \left(\frac{GM}{2\Vinfty^2}\right)^\frac{1}{3}\RG^\frac{2}{3}\left(\left(\frac{\Rc}{\RG}\right)^2+1\right)^\frac{2}{3}
\end{equation}
\par One can also use equation~10 of \citet{Bertin2008} and derive the same formula for the tidal radius.

\par For $\RG\gg\Rc$, the last term on the right-hand side of the \equref{eq:rt} goes to zero and one can approximate the above equation for the logarithmic potential by the following form which is essentially the tidal radius of a system in an isothermal sphere{\footnote{$\rho(r)\propto{r^{-2}}$}} \citep{Baumgardt2003}
\begin{equation}
	\Rt \approx \left(\frac{GM}{2\Vinfty^2}\right)^\frac{1}{3}\RG^\frac{2}{3}, \quad \frac{\RG}{\Rc}\gg1
\end{equation}

\section{Simplifying tidal radii equations for numerical calculations}\label{sec:A2}
It is straightforward to calculate tidal radii using equations~\eqref{eq:rt_lambda} and \eqref{eq:lambda} given an analytical relation for potential. However, when only the matter distribution $\rho(r)$ of the host galaxy is given in analytical form, one needs numerical integration followed by a numerical second order differentiation to find tidal radii which can be a cumbersome and error-prone process.

\par To overcome this issue, we have taken advantage of Poisson's equation, assuming spherical symmetry for the density
\begin{equation}\label{eq:poisson}
\nabla\phi(r)=4\pi G\rho(r) \rightarrow \frac{1}{r^2}\frac{d}{dr}\left(r^2\frac{d\phi(r)}{dr}\right) = 4\pi G \rho(r)
\end{equation}
Expanding this equation yields
\begin{equation}\label{eq:poisson2}
\frac{d^2\phi(r)}{dr^2}+\frac{2}{r}\frac{d\phi(r)}{dr}=4\pi G\rho(r)
\end{equation}
where $d\phi(r)/dr$ is simply the magnitude of acceleration at radius $r$ and can be written in terms of the enclosed mass $M(r)$
\begin{equation}\label{eq:force}
\frac{d\phi(r)}{dr} = G\frac{M(r)}{r^2}
\end{equation}
\par The first derivative of potential with respect to $x_i$ is
\begin{equation}
\frac{\partial\phi(r)}{\partial x_i}=\frac{d\phi(r)}{dr}\frac{\partial r}{\partial x_i}
\end{equation}
knowing that $\partial r/\partial x_i=x_i/r$, then the second derivative of $\phi$ will be
\begin{equation}\label{eq:D2phi}
\frac{\partial^2\phi(r)}{\partial x_i^2}=\frac{d^2\phi(r)}{dr^2}\left(\frac{x_i}{r}\right)^2 + \left(\frac{1}{r} - \frac{x_i^2}{r^3}\right)\frac{d\phi(r)}{dr}
\end{equation}
Substituting \equref{eq:D2phi} into \equref{eq:lambda} yields
\begin{equation}
\lambda_{e,1} = \frac{1}{r}\frac{d\phi(r)}{dr}-\frac{d^2\phi(r)}{dr^2}
\end{equation}
which can be simplified by replacing first and second derivatives of potential with respect to $r$ using equations~\eqref{eq:poisson2} and \eqref{eq:force}
\begin{equation}
\lambda_{e,1} = 3G\frac{M(r)}{r^3} - 4\pi G\rho(r)
\end{equation}
\par Hence the tidal radius of a stellar system with mass $\Mc$, located at distance $\RG$ from the centre of the host galaxy will be 
\begin{equation}\label{eq:rt_num}
\Rt = \Mc^{\frac{1}{3}}\left(3\frac{M(\RG)}{\RG^3}-4\pi\rho(\RG)\right)^{-\frac{1}{3}} = \Mc^{\frac{1}{3}}\left(3\frac{\Omega(\RG)^2}{G}-4\pi\rho(\RG)\right)^{-\frac{1}{3}}
\end{equation}
\noindent where $\Omega(\RG)$ is the orbital angular velocity at $\RG$
$$\Omega(\RG)=\sqrt{G\frac{M(\RG)}{\RG^3}}$$
\par Note that, when $\rho(r)$ is known, \equref{eq:rt_num} only needs one integration, for the enclosed mass of the host galaxy within $\RG$, to calculate the tidal radius.


\bsp
\label{lastpage}
\end{document}